**TITLE: Detection of brain activations induced by naturalistic stimuli in a pseudo model-driven way**

Jiangcong Liu [a, b], Hao Ma [a], Yun Guan [a], Fan Wu [a], Le Xu [a], Yang Zhang [c], Lixia Tian [a, *]

[a] School of Computer and Information Technology, Beijing Jiaotong University, Beijing 100044, China

[b] Beijing Key Laboratory of Traffic Data Analysis and Mining, Beijing Jiaotong University, Beijing 100044, China

[c] Department of Orthopedics, the Seventh Medical Center of Chinese PLA General Hospital, Beijing 100700, China

[*] Corresponding author at: School of Computer and Information Technology, Beijing Jiaotong University, Beijing 100044, China.

E-mail address: lxtian@bjtu.edu.cn (L. Tian)

**Abstract**

Naturalistic fMRI has been suggested to be a powerful alternative for investigations of human brain function. Stimulus-induced activation has been playing an essential role in fMRI-based brain function analyses. Due to the complexity of the stimuli, however, detection of activations induced by naturalistic stimuli (AINSs) has been a tricky problem, as AINS cannot be detected simply in a model-driven way. In this study, we proposed a method to detect AINS in a pseudo model-driven way. Inspired by the strategy of utilizing the commonalities among the brains exposed to the same stimuli for inter-subject correlation analysis, we established response models for one subject by averaging the fMRI signals across several other subjects, and then detected AINSs of the subject using general linear model. We evaluated the effectiveness of AINS with both statistical and predictive analyses on individual differences in sex and intelligence quotient (IQ), based on the four movie fMRI runs included in the Human Connectome Project dataset. The results indicate that AINS is not only sensitive to sex- and IQ-related differences, but also specific enough to decode individuals' sex and IQ. Specifically, activations in brain regions associated with visual-spatial processing were observed to be consistently stronger in the males, and individuals with higher IQ exhibited consistently stronger activations in regions within the visual and the default mode networks. Predictions of individuals' sex and IQ were significantly better than those based on random labels (P < 0.005). Taken together, AINS advanced in this study can be an effective evaluation of human brain function. The conceptual simplicity and easy application of its detection may make AINS a favorable choice for future brain function analyses and personalized medicine based on naturalistic fMRI.

fMRI; prediction; intelligence quotient (IQ).

# 1. Introduction

Functional magnetic resonance imaging (fMRI) has been one of the most popular techniques for studies of human brain function (Biswal et al., 2010; Finn et al., 2015). In addition to the overwhelmingly prevalent task fMRI and resting-state fMRI, naturalistic fMRI has attracted more and more research interests in recent years (Finn, 2021). Using stimuli such as movies, spoken narratives and music for fMRI data acquisition, naturalistic fMRI has been suggested to be able to well approximate real life scenarios and provide an ecologically valid choice for human brain function analyses. Such virtues as lower head motion noise in children and patients with neuropsychiatric diseases and higher subject alertness further support the widespread use of naturalistic fMRI in investigations of human brain function (Eickhoff, Milham, & Vanderwal, 2020; Sonkusare, Breakspear, & Guo, 2019; Vanderwal, Eilbott, & Castellanos, 2015). Regarding its virtues, Sonkusare et al. (2019) even suggest that naturalistic fMRI "may overtake rs-fMRI in clinical applications in the coming years".

A variety of measures have been advanced for brain function analyses based on naturalistic fMRI. Among these, functional connectivity (Benischek et al., 2020), inter-subject correlation (ISC) (Hasson, Nir, Levy, Fuhrmann, & Malach, 2004) and inter-subject functional correlation (Simony et al. 2016) are three of the most influential ones. Studies based on these measures largely improved our understanding of human brain function under real life scenarios (Hasson, Furman, Clark, Dudai, & Davachi, 2008; Nastase, Gazzola, Hasson, & Keysers, 2019), however, these measures focus only on the similarity of fMRI signals. In addition to their *similarity*, the *intensity* of fMRI signals is also informative for brain function analyses. In fact, early fMRI-based studies on human brain function analyzed mainly the changes of fMRI signal intensity induced by abstract stimuli, which have often been referred to as task-induced "activations" (Friston, Frith, Turner, & Frackowiak, 1995; McKiernan, Kaufman, Kucera-Thompson, & Binder, 2003; Stern et al., 1996). There is no doubt that enormous activations would be evoked by naturalistic stimuli, and an effective way to detect such activations will not only benefit investigations of human brain function but facilitate practical applications of naturalistic fMRI (Elliott et al., 2020; Vanderwal, Eilbott, & Castellanos, 2018). To date, however, there is not an effective and convenient method for detection of activations induced by naturalistic stimuli (AINSs).

Methods for activation detection can roughly be classified into two categories, namely, data-driven and model-driven methods (Lee, Zelaya, Amiel, & Brammer, 2008; Zang, Jiang, Lu, He, & Tian, 2004). Such typical data-driven methods as clustering analysis (Filzmoser, Baumgartner, & Moser, 1999; Goutte, Toft, Rostrup, Nielsen, & Hansen, 1999), independent component analysis (McKeown, Hansen, & Sejnowsk, 2003), and principal component analysis (Hansen et al., 1999) have frequently been used for task fMRI data analyses, but none of these methods can provide a direct evaluation of stimulus-induced activations. Model-driven methods have largely been dependent upon the general linear model (GLM), by regressing a response model (often constructed by convolving abstract stimuli with the hemodynamic response function) against fMRI signals (Bullmore et al., 1996; Woolrich, Ripley, Brady, & Smith, 2001; Worsley & Friston, 1995). Being able to provide a direct evaluation of stimulus-induced activations, model-driven methods have been playing a predominant role in activation detection in former task-based fMRI studies.

Due to the high dynamicity and complexity of naturalistic stimuli, model-driven AINS detection has been a tricky problem. Response model construction is of essential importance for model-driven activation detection, and an inappropriate model may lead to a "misinterpretation" of fMRI data (Shahin, Shayegh, Mortaheb, & Amirfattahi, 2016). However, the high dynamicity and complexity of naturalistic stimuli make it impossible for us to construct a response model based on the given stimuli, as is done on abstract stimuli for task-based fMRI studies. Based on the hypothesis that brains exposed to the same naturalistic stimuli would exhibit common spatiotemporal activities, Hasson et al. (2004) used group-average fMRI signals across other subjects for ISC analysis. The strategy of averaging fMRI signals across other subjects can be seen as a roundabout way to model the common responses induced by naturalistic stimuli. Accordingly, a convenient way to construct response models for AINS detection for one subject is to average the fMRI signals across some other subjects, and AINS detection can then be performed in a pseudo model-driven way with the use of GLM.

In this study, we proposed a novel method to detect AINS in a pseudo model-driven way. Specifically, to detect AINS in one region of one subject, we established a response model by averaging the fMRI signals in the region across several other subjects, and then regressed his/her fMRI signal against the response model. We performed the study on the four movie fMRI runs included in the Human Connectome Project (HCP) dataset. To evaluate the effectiveness of AINS we detected, we further performed two sets of statistical analyses to check whether AINS is sensitive to individual differences in sex and intelligence quotient (IQ), and two sets of individualized prediction analyses to check whether AINS is specific enough to decode individuals' sex and IQ. Sex and IQ were taken as test cases in this study for the consideration that sex- and IQ-related differences have long been the focus in the field of neuroscience (Cahill, 2006; Deary, Penke, & Johnson, 2010), and sex classifications and IQ predictions have been much used to test the effectiveness of MRI-derived measures (Genç et al., 2018; Smith et al., 2013), as well as the performance of predictive models (Parisot et al., 2018; Xiao, Stephen, Wilson, Calhoun, & Wang, 2019).

## 2. Materials and methods

### 2.1 Dataset

This study was performed based on the 7 T movie fMRI data included in the HCP S1200 release (https://db.humanconnectome.org/). One hundred and eighty-four healthy young subjects participated in the movie fMRI data acquisition, and four movie fMRI runs were obtained for each subject on two consecutive days. One hundred and seventy-eight subjects (29 ± 3.32 years; 70 males) completed all four runs, and all following analyses were based on the movie fMRI data of them. Written informed consent was obtained from all subjects in accordance with the protocol set by the Human Research Protection Program (Van et al., 2013). Each of the four movie fMRI runs lasted a bit longer than 15 min, and they would be referred to as "Mov1", "Mov2", "Mov3" and "Mov4" according to the acquisition order. The stimuli for Mov1 and Mov3 were composed of short clips (1.0 - 4.1 min) of 4 independent Creative Common videos; those for Mov2 and Mov4 were composed of short clips (3.7 - 4.3 min) of 3 independent Hollywood movies. There was a 20-second black-screen fixation between consecutive clips, before the first clip and after the last clip. At the end of each run, there was a concatenated common repeat clip (1.4 min) for validation purposes. More data details could be found on the HCP website.

We performed the study based on the pre-processed movie fMRI data released on the HCP website. According to the HCP committee, the fMRI data was preprocessed with the Version 3 preprocessing pipelines, which included not only the minimal preprocessing pipelines but also the ICA-FIX sub-pipeline (Glasser et al., 2013). To keep only movie-related fMRI data, volumes corresponding to the 20-second fixations, as well as those corresponding to the short common clip, were removed from each run. For the consideration of hemodynamic response lag, volumes corresponding to the first ten seconds of each clip were removed, and those corresponding to the five seconds right after each clip were kept (Nastase et al., 2019; Simony & Chang, 2020). The remaining movie fMRI data for the four runs included 696, 721, 690 and 704 volumes respectively, which would be the basis for all later analyses.

### 2.2 AINS detection

In this study, we detected AINS in a pseudo model-driven way (but actually data-driven), and ROI-wise AINS detection was performed based on the 264 ROIs defined following the study by Power et al. (2011). Specifically, we first obtained the mean time course of each of the 264 ROIs by averaging the time series across all voxels within the ROI, for each run of each subject. Then the response model for each ROI was obtained by averaging the mean time courses of the ROI across 15 subjects randomly selected from 178 subjects. Finally, the AINS of each ROI of each of the remaining 163 subjects was calculated by regressing his/her time course against the response model of the ROI. To note, we did not establish response models using the leave-one-out strategy (as is the case for ISC calculation) for two considerations: 1) the average fMRI signals across a sub-group of subjects are enough to characterize

the common responses induced by naturalistic stimuli; 2) most importantly, the practice of establishing

response models based on set-aside subjects can help avoid the inter-dependence among the AINS of

different individuals (as is the case for the leave-one-out strategy). We set the number of subjects for

the mean fMRI signal acquisition to 15 for the consideration of obtaining consistent AINSs (see

Section 3.4 for more details), and the same 15 subjects were selected for response model establishment for all four movie fMRI runs. To differentiate from network-wise AINSs below, those for separate ROIs would be referred to as ROI-wise AINSs.

To provide an overall idea about the relative intensity of AINSs in different functional networks,

network-wise AINSs were also obtained. For this purpose, we assigned the 264 ROIs into eleven

functional networks according to the study by Cole et al. (2013), which was also based on the 264

ROIs defined by Power et al. (2011). The eleven networks are the somato-motor network, the cingulo-opercular network, the auditory language network, the default mode network, the visual network, the fronto-parietal network, the salience network, the ventral attention network, the dorsal attention network, the subcortical network and the uncertain network. We calculated the AINS of each network by averaging the ROI-wise AINSs across all ROIs within the network.

*2.3 Statistical analyses of sex- and IQ-related differences in AINS*

To test whether AINS is sensitive to individual differences in apparent traits, we took sex and IQ as test cases and tested sex- and IQ-related differences in AINS. Of the remaining 163 subjects, 66 were male and 97 were female. According to the HCP committee, IQ of the subjects was evaluated using the NIH Cognition Battery Toolbox. Among the 163 remaining subjects, IQ of only 160 subjects were available (91.03 - 153.36, 124.14 $\pm$ 15.15), and all following IQ-related analyses were based on the 160 subjects whose IQ were available.

On each of the 264 ROI-wise AINSs based on each of the four runs, two sample t-test was performed to check whether AINS is sensitive to sex-related differences, and Pearson's correlation between AINS and IQ was performed to check whether AINS is sensitive to IQ-related differences. Consistent significant sex-/IQ-related differences (P < 0.05, uncorrected) based on at least three runs were regarded to be significant. For each of the 2,112 (2 traits $\times$ 4 runs $\times$ 264 ROIs) statistical analyses, 5,000 permutations were performed to check the significance of the differences, and the P-value was calculated as follows:

$$P = \frac{1 + N_{HigherAbsoluteT/R}}{1 + N} \tag{1}$$

where $N$ is the number of permutations, and here $N$ = 5,000; $N_{HigherAbsoluteT/R}$ is the number of permutations in which higher absolute t-scores/r-values were obtained.

In this study, AINS of an ROI would be regarded to be sensitive to sex-/IQ-related differences if at least three of its four t-scores/r-values were significant at P < 0.05 (uncorrected). It is obvious that false positives would less likely to happen if the significant differences occur repeatedly across different runs. To provide a quantitative evaluation of the possibility of false positives in this study, we performed 5,000 random experiments for simulating the chance of observing at least three "positives" by chance. Specifically, in each experiment, we randomly assigned "positive" to an ROI at a possibility of 0.05 and we repeated this random assignment four times (to simulate the four runs in this study). The probability of the ROI exhibiting "positive" in at least three of the four random assignments in 5,000 random experiments can reflect the chance at which we can obtain the current significant differences by chance. According to 5,000 random experiments, it was found that the chance of observing false positives in at least three runs was only 0.0008. That means, though the current threshold of uncorrected P < 0.05 is very loose, a false positive is less likely to occur when we report only sex-/IQ-related differences in at least three runs.

### 2.4 Individualized predictions of sex and IQ based on AINS

To test whether AINS is specific enough to decode individuals' apparent traits, we took sex and IQ as test cases and performed individualized sex classifications and IQ predictions based on AINS. We established predictive models with the widely-used partial least squares regression (Chen, Cao, & Tian, 2019; Yoo et al., 2018), and the number of latent variables was empirically set to 5 (Tian, Ye, Chen, Cao, & Shen, 2021). Predictions were performed on each of the four runs independently, that is, a total of 8 sets (2 tasks × 4 runs) of predictions were performed. For each set of predictions, 10-fold cross-validation was used to evaluate the performance of the predictive models. To avoid possible bias, each set of predictions was repeated 100 times, and the final prediction accuracy was evaluated as the mean accuracy across the 100 rounds. The accuracy of sex classification was evaluated by the percentage of correct classifications, and that of IQ prediction was evaluated by Pearson's correlation between the predicted and actual IQ. We also calculated the area under curve (AUC) and root mean square error (RMSE) of sex classification and IQ prediction, respectively. Five thousand permutations were further performed for each set of predictions to check the significance of the prediction accuracy

(Tian, Ma, & Wang, 2016), and the P-value was calculated as follows (Phipson & Smyth, 2010):

$$P = \frac{1 + N_{HigherAccuracies}}{1 + N} \qquad (2)$$

where $N$ is the number of permutations, and here $N$ = 5,000; $N_{HigherAccuracies}$ is the number of permutations in which the prediction accuracies were observed to be higher than that based on non-permuted sex/IQ.

### 2.5 Analyses of the factors influencing AINS calculation

### 2.5.1 Effects of the number of subjects for mean time series evaluation

Effects of the number of subjects for mean time series evaluations upon AINS analyses were evaluated using intra-class correlation coefficient (ICC). ICC has traditionally been used to evaluate the consistency of scores given by different voters and has been much used to evaluate the consistency of stimulus-induced activations in recent years (Korucuoglu et al., 2020; Schacht et al., 2011). ICC in this study was used to evaluate the consistency of two sets of AINS evaluated based on mean time series extracted using two different groups of subjects, and a high ICC means that the AINSs are stable (less affected by the number of subjects used for response model establishment). We chose to use ICC (1, 1) to keep consistent with most studies on the stability of fMRI-derived measures (Noble, Scheinost, & Constable, 2019), and ROI-wise ICC was calculated for each of the ROI-wise AINSs as follows (Shrout & Fleiss, 1979; Wang et al., 2017):

$$ICC = \frac{MS_b - MS_w}{MS_b + (d-1)*MS_w} \qquad (3)$$

where $MS_b$ represents inter-subject mean square, $MS_w$ represents intra-subject mean square, and d represents the number of observations per subject (d = 2, two group in this study).

Twenty possible subject numbers were considered, and these were 2, 3, 4, 5, 6, 7, 8, 9, 10,11,13, 15, 17, 20, 25, 30, 35, 40, 45 and 50. Based on each possible subject number (e.g., 20 subjects), we randomly selected two groups of subjects (e.g., 20 subjects) for response model establishments. Based on each group of subjects, we then established response models and evaluated AINSs for each of the remaining subjects (138 subjects), as described in Section 2.2. The ROI-wise ICC based on each movie fMRI run was then evaluated based on the AINSs of each ROI (a 138 (subjects) × 2 (response models) matrix), and the global-wise ICC was finally obtained by averaging the ROI-wise ICCs across all 264 ROIs. To avoid possible bias, the experiment on each possible subject number was repeated 10 times, and the final global-wise ICC was evaluated as the mean global-wise ICC across the 10 times. According to Cicchetti and Sparrow (1981), an ICC within the range of ICC < 0.40, $0.40 \leq$ ICC < 0.60, $0.60 \leq$ ICC < 0.75 and $0.75 \leq$ ICC corresponds to "poor", "fair", "good" and "excellent" consistency. In this study, a subject number supporting global-wise ICC $\geq 0.60$ (corresponds to "good" consistency) for all four runs would be regarded to be sufficient for response model establishments.

### 2.5.2 Effects of time series length

Two sets of analyses were carried out to evaluate the effects of time series length upon AINS evaluation. The short common clip included in each of the four movie fMRI runs in the HCP dataset makes it possible for us to evaluate the test-retest reliability of AINS across repeated movie fMRI runs. We performed one set of analyses by first calculating AINSs based on the short common clip included in each of the four movie fMRI runs. Taking a strategy similar to that reported in Section 2.1, we removed the first ten volumes and kept the five volumes right after each clip for the consideration of hemodynamic response lag, and AINSs were evaluated based on the remaining 78 volumes. Again, response models were established based on data of fifteen subjects, and AINSs were evaluated on each of the remaining 163 subjects. ROI-wise ICC was then calculated based on each of the 163 (subjects) × 4 (runs) AINS matrices, and a global-wise ICC was finally obtained by averaging all 264 ROI-wise ICCs.

The other set was performed to test the changes of predictive abilities of AINS with the number of time points for AINS evaluation. Individualized predictions have been much used to evaluate the effectiveness of data preprocessing strategies (Peng, Gong, Beckmann, Vedaldi, & Smith, 2021), fMRI data analyses methods (Lindquist, 2008) and machine learning algorithms (Pereira, Mitchell, & Botvinick, 2009). Here, individualized sex classifications and IQ predictions were utilized to provide a preliminary idea about how the length of movie fMRI time series affects AINS evaluation and ultimately affects prediction accuracies based on it. In this study, twelve possible number of time points were considered, which were 5, 10, 20, 40, 60, 80, 100, 120, 150, 200, 250, 300, 400, 500, and 600. For each possible number of time points, we randomly excerpted a fMRI data segment and repeated all the aforementioned AINS calculation and individualized predictions (in just the same way as those based on whole runs, as described in Sections 2.2 and 2.4). To avoid possible bias caused by random excerption of movie fMRI data segments, the experiment on each possible time point number was repeated 10 times (each based on different starting point of the segment), and the final accuracy was reported as the mean accuracy across the 10 repeats.

(14)

# 3. Results

## 3.1 AINS

Fig. 1 (a) shows the 50 strongest ROI-wise AINSs, and a full list of the ROI-wise AINSs can be found in Table S1 in the Supplemental Material. According to Table S1, enormous difference exists among the ROI-wise AINSs, with some ROIs exhibiting nearly no obvious activations (as indicated by a minimum AINS of 0.013) and some ROIs exhibiting strong activations (as indicated by a maximum AINS of 1.021). The same is true for network-wise AINSs, with the AINSs of the visual network, the ventral attention network, and the dorsal attention network obviously stronger than those of other networks (Fig. 1 (b)).

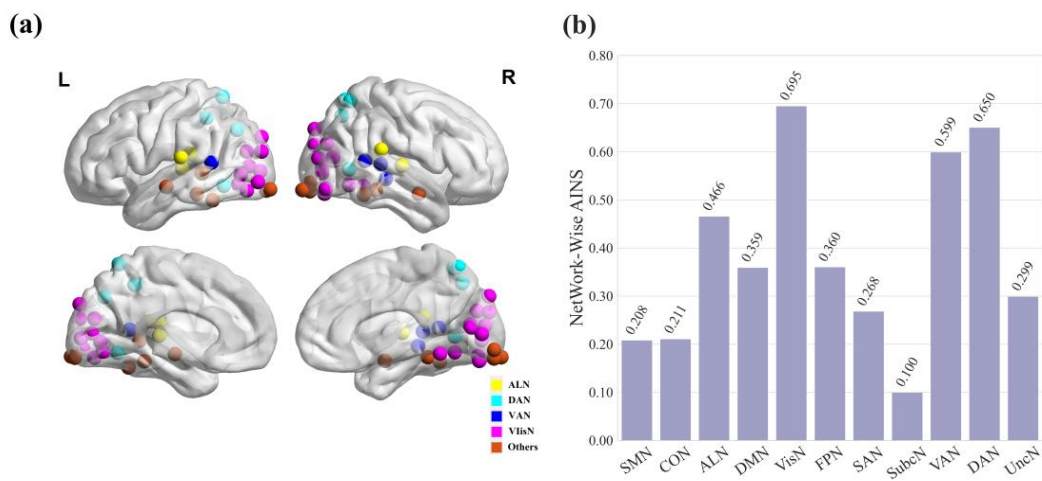

***Fig. 1. Mean ROI- and network-wise AINSs across all runs and all subjects.*** *(a) shows the top fifty strongest AINSs. (b) shows the network-wise AINSs, each calculated by averaging the ROI-wise AINSs within the network. SMN - somato-motor network; CON - cingulo-opercular network; ALN - auditory language network; DMN - default mode network; VisN - visual network; FPN - fronto-parietal network;*

*SAN - salience network; VAN - ventral attention network; DAN - dorsal attention network; SubcN - subcortical network; UncN - uncertain network.*

### 3.2 Sex- and IQ-related differences in AINS

Fig. 2 (a) and Table 1 illustrate the common sex-related AINS differences across runs, and only those showing significant differences based on at least three runs are shown. According to Fig. 2 (a) and Table 1, a majority of the ROIs showing significant sex-related differences lie in the occipital cortex (Brodmann's Area 17, 18 and 19), and two precuneus ROIs also exhibited significant AINS differences. All AINSs exhibiting significant sex-related differences were stronger in the males. The common significant AINS-vs-IQ correlations across runs are shown in Fig. 2 (b) and Table 2. In addition to those within the occipital cortex, significant correlations (P < 0.05, uncorrected) were also observed in such regions as the precuneus (5 ROIs) and the superior temporal gyrus (3 ROIs). The common significant AINS-vs-IQ correlations observed in this study were all positive.

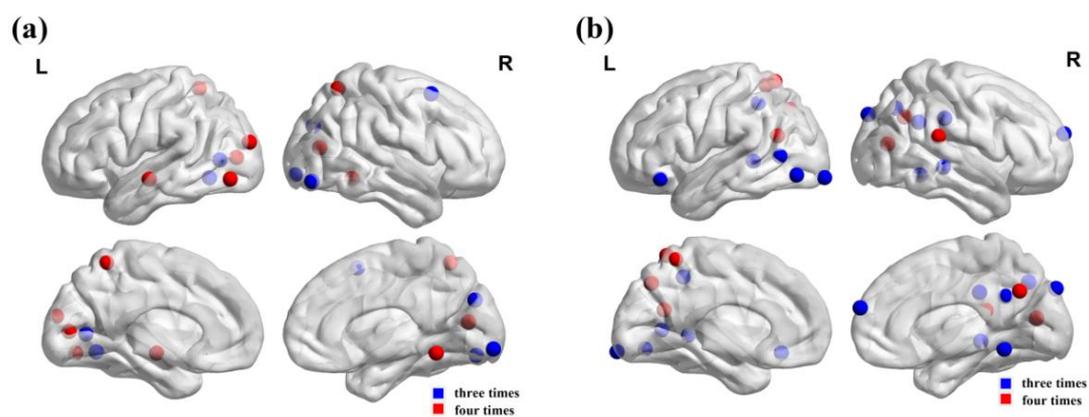

**Fig. 2. Common sex- and IQ-related differences in AINSs across runs.** *(a) and (b) show the common sex- and IQ-related differences. Three/four times denotes the number of the occurrences of significant differences among the four runs. L - left; R - right.*

**Table 1. Common sex-related AINS differences across runs.**

| Number of Occurrences | ROI | MNI Coordinate | Brodmann's Area | Movie No. | t-score (≥) | P-value (≤) |
|---|---|---|---|---|---|---|
| 4 | precuneus | -7,-52,61 | 29 | 1,2,3,4 | 2.893 | 0.004 |
| | lateral occipital cortex | 25,-58,60 | 18 | 1,2,3,4 | 2.580 | 0.009 |
| | occipital pole | -24,-91,19 | 19 | 1,2,3,4 | 2.574 | 0.024 |
| | lateral occipital cortex | 40,-72,14 | 18 | 1,2,3,4 | 2.552 | 0.010 |
| | intracalcarine cortex | -8,-81,7 | 17 | 1,2,3,4 | 2.139 | 0.028 |
| | lateral occipital cortex | -47,-76,-10 | 18 | 1,2,3,4 | 2.125 | 0.027 |
| | lingual gyrus | 18,-47,-10 | 17 | 1,2,3,4 | 2.016 | 0.038 |
| | central opercular cortex | -56,-13,-10 | 44 | 1,2,3,4 | 1.996 | 0.044 |
| 3 | intracalcarine cortex | -18,-68,5 | 17 | 1,2,3 | 2.366 | 0.020 |
| | inferior temporal gyrus | -42,-60,-9 | 20 | 2,3,4 | 2.323 | 0.032 |
| | lateral occipital cortex | 43,-78,-12 | 18 | 1,3,4 | 2.275 | 0.021 |
| | precuneus | 15,-77,31 | 29 | 1,2,3 | 2.238 | 0.022 |
| | middle frontal gyrus | 32,14,56 | 10 | 1,2,3 | 2.222 | 0.028 |
| | occipital pole | 8,-91,-7 | 19 | 1,2,3 | 2.080 | 0.032 |

*t-score ( ≥ ) denotes the minimum t-score among the four/three runs; P-value ( ≤ ) denotes the maximum P-value among the four/three runs.*

**Table 2. Significant correlation between IQ and AINS across runs.**

| Number of Occurrences | ROI | MNI Coordinate | Brodmann's Area | Movie No. | r-value (≥) | P-value (≤) |
|---|---|---|---|---|---|---|
| 4 | precuneus | -7,-52,61 | 29 | 1,2,3,4 | 0.323 | 0.0002 |
| | superior temporal gyrus | 65,-33,20 | 38 | 1,2,3,4 | 0.242 | 0.002 |
| | precuneus | 6,-59,35 | 29 | 1,2,3,4 | 0.205 | 0.006 |
| | lateral occipital cortex | 40,-72,14 | 18 | 1,2,3,4 | 0.200 | 0.006 |
| | angular gyrus | -46,-61,21 | 39 | 1,2,3,4 | 0.195 | 0.008 |
| | lateral occipital cortex | -17,-59,64 | 18 | 1,2,3,4 | 0.170 | 0.015 |
| | precuneus | -7,-71,42 | 29 | 1,2,3,4 | 0.162 | 0.020 |
| 3 | superior temporal gyrus | 51,-29,-4 | 38 | 1,2,3 | 0.212 | 0.003 |
| | superior temporal gyrus | -49,-42,1 | 38 | 1,2,4 | 0.204 | 0.004 |
| | frontal pole | 6,64,22 | 10 | 1,2,4 | 0.198 | 0.007 |
| | superior parietal lobule | -33,-46,47 | 7 | 1,2,3 | 0.192 | 0.007 |

| | occipital pole | 15,-87,37 | 19 | 1,2,3 | 0.188 | 0.010 |
|---|---|---|---|---|---|---|
| | precuneus | 8,-48,31 | 29 | 1,2,3 | 0.185 | 0.010 |
| | parietal operculu cortex | 54,-28,34 | 43 | 1,2,3 | 0.184 | 0.012 |
| | lateral occipital cortex | -52,-63,5 | 18 | 1,3,4 | 0.174 | 0.014 |
| | lingual gyrus | 18,-47,-10 | 17 | 1,2,3 | 0.171 | 0.014 |
| | lateral occipital cortex | -47,-76,-10 | 18 | 1,2,3 | 0.165 | 0.020 |
| | frontal pole | -46,31,-13 | 10 | 1,2,3 | 0.164 | 0.016 |
| | precuneus | 11,-66,42 | 29 | 1,2,4 | 0.160 | 0.022 |
| | occipital pole | -25,-98,-12 | 19 | 1,2,4 | 0.158 | 0.024 |

*r-value ( ≥ ) denotes the minimum r-value among the four/three runs; P-value ( ≤ ) denotes the maximum P-value among the four/three runs.*

### 3.3 Prediction accuracies of sex and IQ based on AINSs

With sex classification and IQ prediction used as test cases, we evaluated the ability of AINS in predictions of individuals' apparent traits. The sex classification accuracies based on the four runs were 65.5%, 74.2%, 67.5% and 73.8%, respectively (Fig. 3 (a)). Though these accuracies were far below perfect classification, among the 5,000 permutations performed for each set of sex classifications, 4, 0, 0 and 0 higher accuracies were observed (Fig. 3 (b-e)), corresponding to P-values of 0.0010, 0.0002, 0.0002 and 0.0002, respectively. The correlation between the actual and predicted IQ based on the four runs were 0.317, 0.245, 0.362 and 0.322, respectively (Fig. 3 (f)). Among the 5,000 permutations performed for each set of IQ predictions, 1, 15, 0 and 0 better predictions were observed (Fig. 3 (g-j)), corresponding to P-values of 0.0004, 0.0032, 0.0002 and 0.0002, respectively. Fig. 4 (a-d) were the representative ROC curves of the sex classifications based on each of the four runs, respectively; Fig. 4 (e-h) were representative plots of predicted-vs-actual IQ based on each of the four runs, respectively.

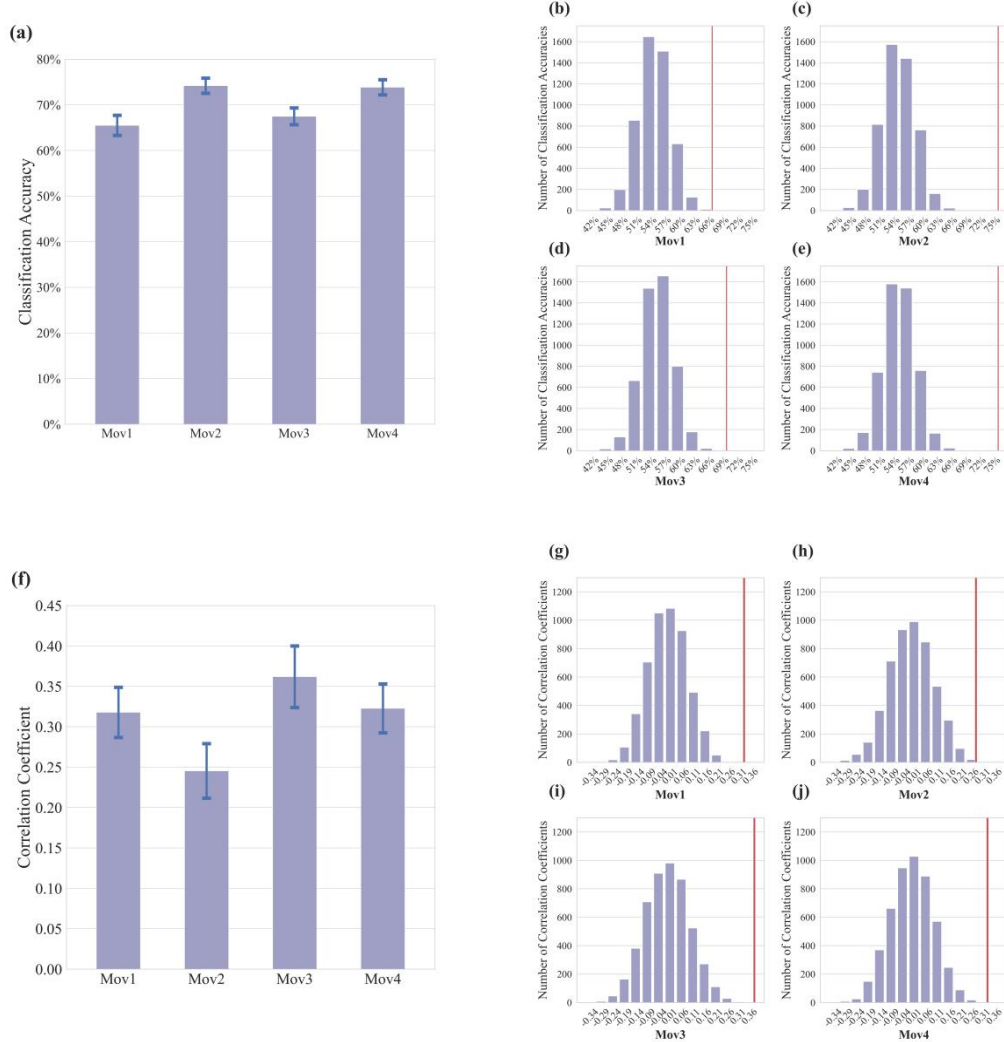

**Fig. 3. Performance of predictive models based on AINSs.** *(a) shows the accuracies of sex classifications, and (f) shows Pearson's correlations between the actual and predicted IQ. The error bars in (a) and (f) indicate standard deviation of accuracies across 100 rounds of predictions. (b, c, d, e) and (g, h, i, j) are the distribution of accuracies for sex classifications and IQ predictions based on permutation tests, and the red vertical lines in them represent the accuracies based on actual sex /IQ. "Movi" indicates results based on the i-th movie fMRI run.*

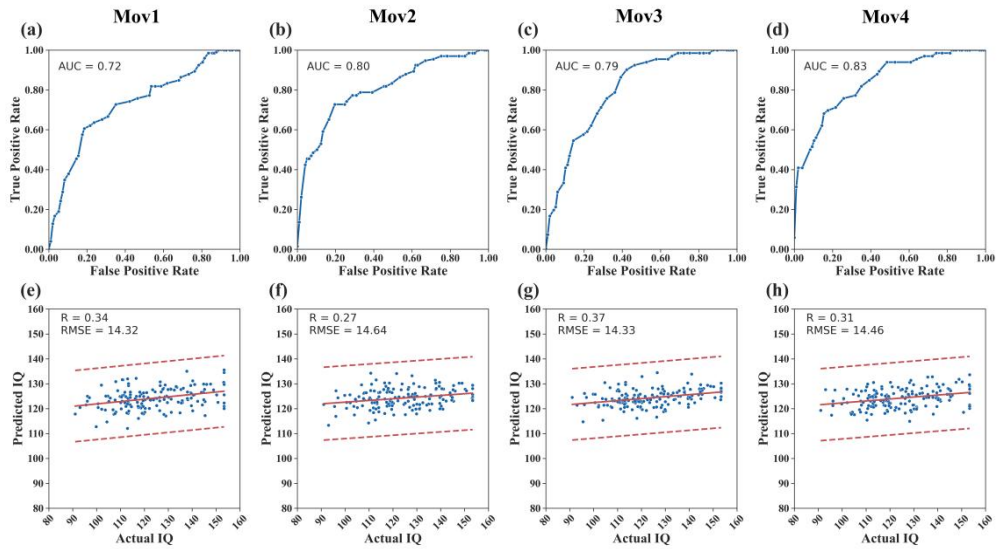

**Fig. 4. Representative ROC curves of sex classifications (a-d) and representative plots of predicted-vs-actual IQ (e-h).** *"Movi" indicates sex classification/IQ prediction based on the i-th movie fMRI run.*

### 3.4 Factors influencing AINS calculation

Fig. 5 illustrates the changes of global-wise ICC of AINSs with the number of subjects used for response model establishments. According to Fig. 5, the global-wise ICC of AINSs increased progressively with the number of subjects used for response model establishments. AINSs based on models established using 6, 15 and 40 subjects were found to be of "fair" (0.60 > ICC ≥ 0.4), "good" (0.75 > ICC ≥ 0.60) and "excellent" (ICC ≥ 0.75) consistency for all four runs, respectively (Cicchetti & Sparrow, 1981). These results indicate that AINS detection requires at least 6 (preferably 15) subjects for response model establishments.

The global-wise ICC of AINSs based on the repeated movie fMRI clip was only 0.156, and this indicates "poor" test-retest reliability of AINSs based on short fMRI runs. Fig. 6 shows the changes of the accuracies of predictive models with the number of time points used for AINS calculation. According to Fig. 6, the performance of predictive models improved progressively with the number of time points for AINS calculation, and this performance improvement can be fitted well with a log function ($R^2 = 0.595$ for sex classification, and $R^2 = 0.415$ for IQ prediction) (Fig. 6 (c) and (d)). Taken together, these results demonstrate that longer movie fMRI runs are favorable for AINS evaluation.

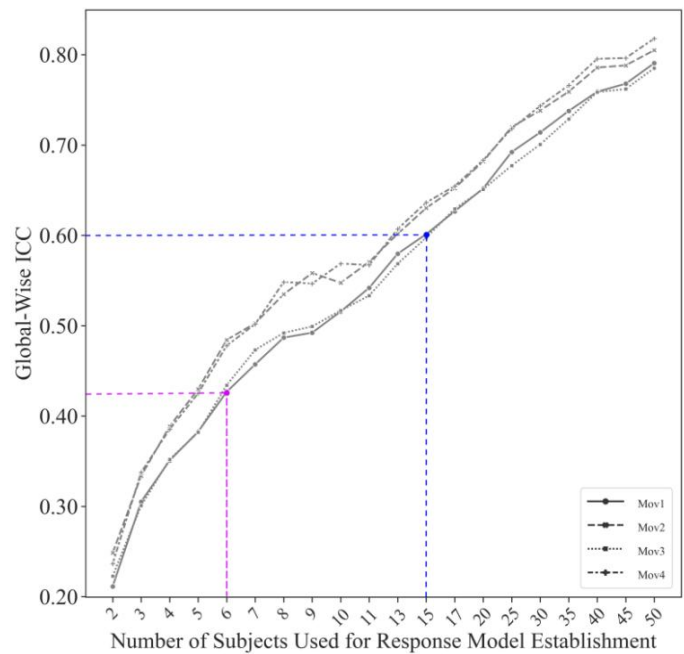

**Fig. 5. Changes of global-wise ICC of AINSs with the number of subjects used for response model establishments.** *ICC was used to evaluate the consistency of two sets of AINS, with each set calculated based on response models established using a separate group of subjects. Both ROI- and global-wise ICC were evaluated based on each movie fMRI run separately.*

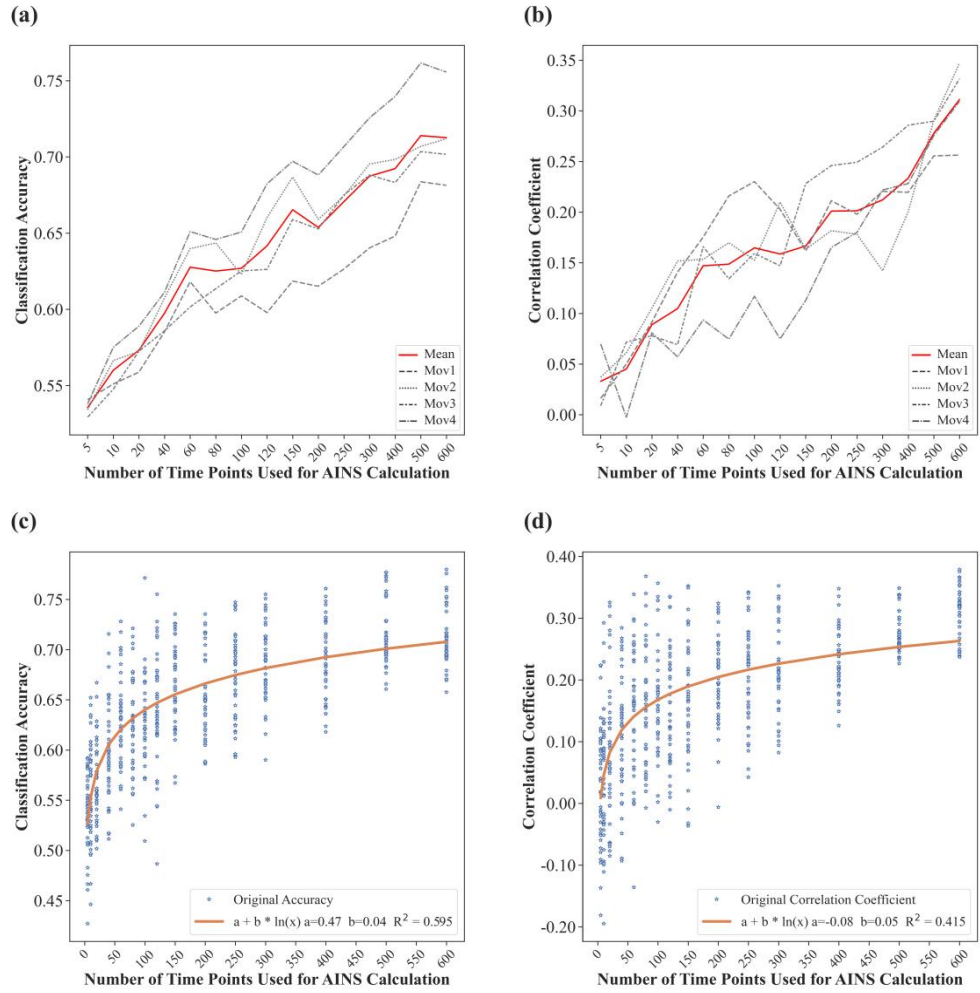

**Fig. 6. Changes of the performance of predictive models with the number of time points used for AINS calculation.** *(a) and (b) show the changes of the accuracies of sex classifications and IQ predictions with the number of time points. "Mean" denotes the average accuracy across the four runs based on the given number of time points. (c) and (d) show the fits of the predictive model performance with log functions. It can be seen that the performance of the predictive model can be well fitted by a log function of the number of time points used for AINS calculation.*

## 4. Discussion

In this study, we advanced a novel method to detect AINS in a pseudo model-driven way. Based on the commonalities among the brains exposed to the same stimuli, we established the response model by averaging the fMRI signals in a region across several other subjects, and then detected the activation in the region of one subject by regressing his/her fMRI signal against the mean signal. We demonstrated that AINSs were sensitive to individual differences in apparent traits (Fig. 2) by two sets of statistical analyses. Furthermore, individual differences in AINS were found to be specific enough to decode the apparent traits (Figs. 3 and 4). These findings demonstrated the potential of AINS in future studies on human brain function and precise and personalized medicine. The effects of the number of time series lengths and the number of subjects for response model establishments upon AINS evaluation were also analyzed (Figs. 5 and 6). The following are detailed discussions over the results.

### 4.1  About AINS

The complexity of naturalistic stimuli makes it impossible to establish a response model for AINS detection simply by convolving the external stimuli with the hemodynamic response function, as is performed for activation detection based on task fMRI. In this study, we established response models for one subject by averaging the fMRI signals across several other subjects to simulate the common spatiotemporal activities and detected AINS by regressing his/her fMRI signal against the mean signal. The strategy of averaging the fMRI signals across several subjects can enhance the signal-to-noise ratio and preclude stimuli-irrelevant signals (e.g., scanner or physiological noise) (Dubois & Adolphs, 2016; Simony et al., 2016). Like the response model established based on hemodynamic response function for classic task activation detection, the response model established here would largely conceal individuals' specific temporal activation patterns. Though there have been several recent activation detection strategies which may also be applicable to naturalistic fMRI data (Dong et al., 2020; Hütel,

Melbourne, & Ourselin, 2018), the complexity of the detection models/strategies largely hamper their practical applications. The conceptual simplicity and easy application of AINS detection may make it a favorable choice for future naturalistic fMRI data analyses.

Longer fMRI signals and more subjects used for response model establishments were found to be favorable for AINS detection (Figs. 5 and 6). Naturalistic stimuli are more acceptable for subjects (Finn et al., 2020; Sonkusare et al., 2019), and the requirement of longer fMRI signals would accordingly not hamper future applications of AINS. According to Fig. 5, the response models established on the 15 subjects can produce consistent AINSs (global-wise ICC ≥ 0.6), and response models based on only 6 subjects could produce AINSs of "fair" consistency (global-wise ICC ≥ 0.4). To note, even this consistency (global-wise ICC ≥ 0.4) was higher than that of the overwhelmingly prevalent resting-state functional connectivity (ICC = 0.29, according to the meta-analysis of Noble et al. (2019)). In future applications, response models could in fact be established in a leave-one-out way (Hasson et al., 2004; Simony et al., 2016), with the response model of one ROI for one subject established by averaging the fMRI signals of the ROI across all other subjects, if cross-validation is not involved in.

Among the eleven networks, the visual network, the ventral attention network, the dorsal attention network and the auditory language network exhibited relatively stronger AINS, and the subcortical network exhibited the weakest AINS (Fig. 1). The visual network supports mainly visual perception and visual information processing, such as face recognition and object tracking (Kay, Naselaris, Prenger, & Gallant, 2008; Tootell et al., 1997); The dorsal attention network supports goal-directed top-down attentional control (Giesbrecht, Woldorff, Song, & Mangun, 2003; Hopfinger, Buonocore, &

Mangun, 2000) and the ventral attention network supports stimulus-driven bottom-up attentional reorienting (Astafiev et al., 2003; Kincade, Abrams, Astafiev, Shulman, & Corbetta, 2005); The auditory language network supports auditory and language processing (Bookheimer, 2002; Thivard et al., 2005). Each of these functions is essential for movie watching, and the results of stronger AINSs in the four networks are consistent with the fact that movie watching entails more involvement of them. The subcortical network has traditionally been regarded to support such low-level functions as autonomic control and nociception (Gianaros, Van der Veen, & Jennings, 2004; Constant, & Sabourdin, 2015), and the present finding of the weakest AINS in the subcortical network is consistent with the fact that movie watching requires little autonomic control.

### 4.2  Potential applications of AINSs

Sex-related AINS differences were mostly found in the occipital regions (Brodmanns' Area 17, 18 and 19), with 6 of 8 ROIs exhibiting significant sex-related differences across all four runs lie in the occipital lobe (Fig. 2 (a) and Table 1). AINSs of these ROIs were uniformly stronger in the males, and we suggest that this finding is consistent with the advantages of the males in visuospatial information processing (Scholten, Aleman, & Kahn, 2008; Sommer, Aleman, Bouma, & Kahn, 2004; Sowell et al., 2007). Specifically, the occipital regions have traditionally been linked to visual perception and visual information processing (Sowell et al., 2007), and the superiority of the males in visual-spatial tasks have been much associated with their stronger activations in the occipital regions (Scholten et al., 2008; Sommer et al., 2004).

In addition to those in the occipital cortex, quite a few ROIs at the precuneus and the superior temporal cortex exhibited significant positive AINS-vs-IQ correlations (Fig. 2 (b) and Table 2). The precuneus is a critical component of the posterior default mode network, which has been suggested to be an active and dynamic 'sense-making' network that integrates extrinsic information with prior intrinsic information to form rich, context dependent models of situations (Yeshurun, Nguyen, & Hasson, 2021). The precuneus itself has been reported to be essential for integrating sensory and motor information to represent word concepts (Fernandino et al., 2016) and speaker-listener coupling (Stephens, Silbert, & Hasson, 2010; Silbert, Honey, Simony, Poeppel, & Hasson, 2014). The superior temporal cortex plays critical roles in hearing, speech, and language (for a review, see Karnath, 2001), and each of the functions is essential for movie watching. Taken together, the present finding of significant AINS-vs-IQ correlations in the occipital regions, as well as those in the precuneus and the superior temporal cortex, is consistent with the essential roles of the regions in dealing with incoming extrinsic information (visual and auditory information here) and integrating the extrinsic and prior intrinsic information, that is, individuals with higher IQ may be more efficient in integrating extrinsic and prior intrinsic information and accordingly adapt better to the environments.

Besides its sensitivity to sex- and IQ-related differences, AINS was also found to be specific enough to decode individuals' sex and IQ (Figs. 3 and 4). Though performed on healthy subjects, its predictive ability here indicates the potential of AINS in clinical applications. In combination with the virtues of naturalistic fMRI such as relatively less contamination by head motion noise, more controlled nature and more ecological validity (Sonkusare et al., 2019), AINS can be a favorable choice for future clinical applications. Though both sex and IQ can be predicted with AINS extracted using each of the four movie fMRI runs, it should be noted that the prediction accuracies varied much, with sex predicted better based on Mov2 and Mov4, and IQ predicted the best based on Mov3 but the worst based on Mov2 (Fig. 3). That is, the performance of a predictive model based on AINS is dependent upon the given naturalistic stimuli, and utilizing appropriate stimuli can effectively promote its practical usage. Further studies are expected to figure out the influences of naturalistic stimuli upon the performance of predictive models before the wide practical applications of AINS.

### 4.3 Methodological issues

Three methodological issues should be addressed. Firstly, individual differences in head motion may confound AINS calculation. To evaluate the influences of head motion on AINS, we directly correlated each of the ROI-wise AINS with the mean frame-wise displacements of subjects. The mean correlations (across 264 ROIs) were only 0.011, -0.007, 0.009 and 0.026 for the four runs respectively, which are by far lower than a significant correlation (R = 0.154, P = 0.05, degree of freedom = 161). That is, head motion may have only limited influence upon the current results. Secondly, we performed activation detection based only on movie fMRI data in this study. Further studies are expected to evaluate AINS based on other types of naturalistic stimuli, e.g., story narration and music. Finally, the concatenations of movie clips were used as stimuli in this study, and further studies using fMRI data based on continuous stimuli may provide a better understanding of the abilities of AINS.

### 5.   Conclusions

Activation induced by external stimuli has traditionally been an effective evaluation of human brain function, while the complexity of the naturalistic stimuli makes AINS detection a tricky problem. In this study, we proposed a novel method to detect AINS in a pseudo model-driven way, through establishing a response model for one subject with the average fMRI signal across other subjects and then detecting AINS with the use of GLM. According to the statistical and predictive analyses, AINS was not only sensitive to sex- and IQ-related individual differences, but specific enough to decode sex and IQ. The conceptual simplicity and easy application of AINS detection, together with the virtues of naturalistic fMRI, may make AINS a favorable choice for future brain function analyses and personalized medicine.

**Acknowledgments**

This work was supported by the National Natural Science Foundation of China (Grant Nos. 62276021, 61773048). Data were provided by the Human Connectome Project, WU-Minn Consortium (Principal Investigators: David Van Essen and Kamil Ugurbil; 1U54MH091657) funded by the 16 NIH Institutes and Centers that support the NIH Blueprint for Neuroscience Research; and by the McDonnell Center for Systems Neuroscience at Washington University. The funding sources had no involvement in the preparation of the paper.

**Declaration of Competing Interest**

None.

**Funding**

This work was supported by the National Natural Science Foundation of China (Grant Nos. 62276021, 61773048).

**Data and code availability**

Data used in this study were provided by the Human Connectome Project, WU-Minn Consortium (Principal Investigators: David Van Essen and Kamil Ugurbil; 1U54MH091657). The Human Connectome Project data are publicly available at http://www.humanconnectomeproject.org/data/. The code are available for download from https://github.com/tianbjtu/AINS.

**CRediT authorship contribution statement**

Jiangcong Liu: Conceptualization, Formal analysis, Investigation, Methodology, Visualization, Writing - original draft, Writing - review & editing; Hao Ma: Formal analysis, Investigation, Writing - review & editing; Yun Guan: Investigation, Writing - review & editing; Fan Wu: Investigation, Writing - review & editing; Le Xu: Investigation, Writing - review & editing; Yang Zhang: Investigation, Writing - review & editing; Lixia Tian: Conceptualization, Data curation, Funding acquisition, Investigation, Methodology, Writing - review & editing.